\documentclass[aps,twocolumn,superscriptaddress]{revtex4}
\usepackage{epsfig}
\usepackage{times}
\usepackage{color}
\begin{document}

\title{Grand challenges in social physics: In pursuit of moral behavior}

\author{Valerio Capraro}
\email{v.capraro@mdx.ac.uk}
\affiliation{Department of Economics, Middlesex University, The Burroughs, London NW4 4BT, U.K.}

\author{Matja{\v z} Perc}
\email{matjaz.perc@gmail.com}
\affiliation{Faculty of Natural Sciences and Mathematics, University of Maribor, Koro{\v s}ka cesta 160, SI-2000 Maribor, Slovenia}
\affiliation{Complexity Science Hub Vienna, Josefst{\"a}dterstra{\ss}e 39, A-1080 Vienna, Austria}

\begin{abstract}
Methods of statistical physics have proven valuable for studying the evolution of cooperation in social dilemma games. However, recent empirical research shows that cooperative behavior in social dilemmas is only one kind of a more general class of behavior, namely moral behavior, which includes reciprocity, respecting others' property, honesty, equity, efficiency, as well as many others. Inspired by these experimental works, we here open up the path towards studying other forms of moral behavior with methods of statistical physics. We argue that this is a far-reaching direction for future research that can help us answer fundamental questions about human sociality. Why did our societies evolve as they did? What moral principles are more likely to emerge? What happens when different moral principles clash? Can we predict the break out of moral conflicts in advance and contribute to their solution? These are amongst the most important questions of our time, and methods of statistical physics could lead to new insights and contribute towards finding answers.
\end{abstract}

\maketitle

\section*{Introduction}

Our time now is unique and special in that we are arguably richer, safer, and healthier than ever before \cite{pinker2011better, pinker2018enlightenment}, but simultaneously, we are also facing some of the greatest challenges of our evolution. Climate change, the depletion of natural resources, staggering inequality, the spread of misinformation, persistent armed conflicts, just to name a few examples, all require our best efforts to act together and to renounce part of our individual interests for the greater good. Understanding when, why, and how people deviate from their best self-interest to act pro-socially, benefitting other people and the society as a whole, is thus amongst the most important aims of contemporary scientific research.

Pro-social behavior can come in many forms, the most studied of which is cooperation. Indeed, cooperation is so important that many have contended that our capacity to cooperate at large scales with unrelated others is what makes human societies so successful \cite{ostrom2000collective, fehr2002altruistic, milinski2002reputation, gintis2003explaining, fehr2004social, henrich2006costly, nowak2006five, herrmann2008antisocial, bowles2011cooperative, capraro2013model, rand2013human, perc2017statistical}. Moreover, the psychological basis of cooperation, \emph{shared intentionality}, that is, `the ability and motivation to engage with others in collaborative, co-operative activities with joint goals and intentions' is what makes humans \emph{uniquely} human, as it is possessed by children, but not by great apes \cite{tomasello2005search}.

Although human cooperation is believed to originate from our evolutionary struggles for survival \cite{hrdy_11}, it is clear that the challenges that pressured our ancestors into cooperation today are gone. Nevertheless, we are still cooperating, and on ever larger scales, to the point that we may deserve being called `SuperCooperators' \cite{nowak_11}. Taking nothing away from the immense importance of cooperation for our evolutionary success and for the wellbeing of our societies, recent empirical research shows, however, that to cooperate is just a particular manifestation of moral behavior \cite{capraro2018right}. And while methods of statistical physics have been used prolifically to study cooperation \cite{perc2017statistical}, other forms of moral behavior have not. Our goal here is to outline the many possibilities for future research at the interface between physics and moral behavior, beyond the traditional framework of cooperation in social dilemmas.

\section*{Cooperation}

To study cooperative behavior, scientists use social dilemma games, such as the prisoner's dilemma \cite{rapoport1965prisoner}, the stag hunt \cite{skyrms_04}, or the public goods game \cite{hardin2009tragedy}. In these games, players have to decide whether to cooperate or to defect: cooperation maximizes the payoff of the group, while defection maximizes the payoff of an individual. This leads to a conflict between individual and group interests, which is at the heart of each social dilemma, and in particular at the heart of the cooperation problem.

Since cooperating is not individually optimal, cooperative behavior cannot evolve among self-interested individuals, unless other mechanisms are at play. Several mechanisms for the evolution of cooperation have been identified and studied, including kin selection \cite{hamilton1964genetical}, direct reciprocity \cite{trivers1971evolution}, indirect reciprocity \cite{nowak1998evolution}, social preferences \cite{fehr1999theory, bolton2000erc, charness2002understanding, engelmann2004inequality}, the internalization of social heuristics \cite{rand2012spontaneous}, translucency \cite{capraro2015translucent}, cooperative equilibria \cite{halpern2010cooperative, capraro2013model, barcelo2015group}, as well as many others.

One realistic mechanism for the evolution of cooperation is network reciprocity. Everyday interactions among humans do not happen in a vacuum. We are more likely to interact and cooperate within our network of family members, friends, and coworkers, and we rarely interact, let alone cooperate, with strangers. One can formalize this situation by assuming that individuals occupy the vertices of a graph and interact only with their neighbors. Can this spatial structure promote the evolution of cooperation? The answer is yes \cite{rand_pnas14}. And the intuition is that, in this setting, cooperators can form clusters and protect themselves from the invasion of defectors \cite{nowak1992evolutionary, lieberman2005evolutionary, ohtsuki2006simple}. These `games on graphs are difficult to analyze mathematically', but `are easy to study by computer simulations' \cite{nowak2006five}. Games on networks present the natural setting in which one can apply the techniques and methods of statistical physics and network science to study cooperation \cite{perc_bs10, wang_epjb15}, as well as other forms of moral behavior.

\section*{Statistical physics of human cooperation}

Methods of statistical physics have come a long way in improving our understanding of the emergence of cooperation and the phase transitions leading to other counterintuitive evolutionary outcomes. Research has revealed that such outcomes depend on the structure of the social network, the type and strength of interactions, and on the complexity and number of competing strategies \cite{santos_prl05, pacheco_prl06, gomez-gardenes_prl07, ohtsuki_prl07, lee_s_prl11,  mathiesen_prl11, szolnoki2012defense, assaf_prl12, gomez_prl13, knebel_prl13, pinheiro_prl14}. Aspects particularly relevant to human cooperation have also been studied in much detail \cite{perc2017statistical}. The workhorse behind this research has been the spatial public goods game \cite{szolnoki_pre09c, perc_ejp17}, with extensions towards different forms of punishment \cite{helbing_njp10, szolnoki_pre11a, perc_njp12, wang2013impact, chen_njp14, chen_pre15, szolnoki_prx17}, rewarding \cite{szolnoki_epl10, hilbe_prsb10, szolnoki_njp12, szolnoki_prsb15}, and tolerance \cite{szolnoki_njp16}, to name just some examples. The Monte Carlo method is thereby typically used \cite{perc_ejp18}, which ensures that the research is aligned with statistical physics methodology. This in turn enables a comparison of simulation results with generalized mean-field approximations \cite{dickman_pre01, szolnoki_pre02, dickman_pre02}, and a proper determination of phase transitions between different stable strategy configurations \cite{szolnoki2012defense}. Ultimately, the goal is to identify and understand pattern formation, the spatiotemporal dynamics of solutions, and the principles of self-organization that may lead to socially favorable evolutionary outcomes.

As an example of an evolutionary  game that yields an impressively intricate phase diagram, we mention an $8$-strategy public goods game with diverse tolerance levels. The phase diagram is presented in Fig.~1 of \cite{szolnoki_njp16}, based on which several observations can be made. In the first place, we can  observe that higher tolerance levels are supported by higher multiplication factors in the public goods game, and vice versa. This is in agreement with experience, in that overly tolerant strategies cannot survive in the presence of other less tolerant strategies. From the viewpoint of the considered $8$-strategy public goods game this is also not surprising, because players adopting the most tolerant strategy act as loners only if everybody else in the group is a defector. And evidently, such naive tolerance can not compete with other less tolerant strategies that also compete in the game. Secondly, the phase diagram also reveals that if the cost of inspection is too high, or if the value of the multiplication factor is either very low or very high, tolerance is not viable at all. Even if one tries out different tolerance levels, the evolutionary pressure from other strategies is simply too strong.

While it is beyond the scope of this work to go further into details, it should be noted that phase diagrams as the one presented in Fig.~1 of \cite{szolnoki_njp16} provide an in-depth understanding of the evolutionary dynamics and of the phase transitions that lead from one stable strategy configuration to the other. The key for obtaining accurate locations of phase transition points and the correct phases is the application of the stability analysis of competing subsystem solutions \cite{perc_ejp18}. A subsystem solution can be formed by any subset of all the competing strategies, and on their own (if separated from other strategies) these subsystems solutions are stable. If the subsystem solution is formed by a single strategy this is trivially true, but it can be likewise true if more than one strategy forms a subsystem solution. The dominant subsystem solution, and hence the phase that is ultimately depicted in the phase diagram as the stable solution of the whole system, can only be determined by letting all the subsystem solutions compete against each other.

By means of this approach, several important insights have been obtained. By peer-based strategies, for example, we note the importance of indirect territorial competition in peer punishment \cite{helbing_ploscb10}, the spontaneous emergence of cyclic dominance in rewarding \cite{szolnoki_epl10}, and an exotic first-order phase transition observed with correlated strategies \cite{szolnoki_prx13}. By institutionalized strategies, we have observed the fascinating spatiotemporal complexity that is due to pool punishment \cite{szolnoki_pre11a}, while in the realm of self-organization of incentives for cooperation, we have demonstrated the elevated effectiveness of adaptive punishment \cite{perc_njp12}, the possibility of probabilistic sharing to solve the problem of costly punishment \cite{chen_njp14}, and the many evolutionary advantages of adaptive rewarding \cite{szolnoki_njp12}. With antisocial strategies, we have shown the restoration of the effectiveness of prosocial punishment when accounting for second-order free-riding on antisocial punishment \cite{szolnoki_prx17}, and the rather surprising lack of adverse effects with antisocial rewarding \cite{szolnoki_prsb15}.

While this is just a short snippet of statistical physics research concerning human cooperation, it hopefully showcases successfully the potency of the approach for studying complex mathematical models that describe human behavior, thus recommending itself also for relevantly studying other types of moral behavior to which we attend to in what follows.

\section*{Moral behavior}

Empirical research has indeed shown  that cooperation in social dilemmas is only one facet of a more general class of behavior -- moral behavior. When subjects are asked to report what they think is the morally right thing to do in social dilemmas, they typically answer: `to cooperate' \cite{capraro2018right}.

Morality is universal across human societies. Virtually all societies adopt behavioral rules that are presented to the people as moral principles. But where do these rules come from? A classical non-scientific explanation, still adopted by many societies and religious thinkers, is that they are emanated directly from God. However, in recent years, social scientists have been developing a scientific theory of morality, according to which morality evolved as a a mechanism `to promote and sustain cooperation' \cite{greene2015rise}. As psychology-star Michael Tomasello put it: `human morality arose evolutionarily as a set of skills and motives for cooperating with others' \cite{tomasello2013origins}. Similar positions have also been put forward in \cite{rawls2009theory, mackie1990ethics, wong1984moral, rai2011moral, curry2016morality, curry2018good}. However, the word `cooperation' in these statements does not refer only to cooperation in social dilemmas. How does this general form of cooperation translates into specific behaviors?

A recent study exploring morality in $60$ societies across the world found that seven moral rules are universal: love your family, help your group, return favors, be brave, defer to authority, be fair, and respect others' property. Although, what is not universal is how they are ranked \cite{curry2018good}. Of course, not all these rules are easy to study using simple games on networks, but some are. For example, `returning favors' can be studied using a sequential prisoner's dilemma, in which the players do not choose their strategy simultaneously, but sequentially. Alternatively, it can be studied using the trust game \cite{berg1995trust}. In the trust game, player $1$ starts with a sum of money and has to decide how much of it, if any, to transfer to player 2. Any amount transferred gets multiplied by a factor larger than 1 and handed to player 2. Then player 2 has to decide how much of it to keep and how much of it to return to player 1.

Similarly, `help your group' can be studied using games with labeled players, in which agents come with a label representing the group(s) they belong to \cite{tajfel1970experiments}. `Fairness' can be studied using the ultimatum game \cite{guth1982experimental}, as has already been done along these lines \cite{page_prsb00, kuperman_epjb08, eguiluz_acs09, da-silva_r_jtb09, deng_ll_pa11, gao_j_epl11, szolnoki2012defense, szolnoki2012accuracy, deng_ll_jsm12, iranzo_pone12, miyaji_csf13}, or the dictator game \cite{forsythe1994fairness}. `Respect others' property' can be studied using games with special frames, as, for example, the Dictator game in the Take frame, for which it is known that taking is considered to be more morally wrong than giving \cite{krupka2013identifying}.

Beyond these seven rules, there are other forms of moral behavior that are worth studying, as, for example, `honesty'. A common game theoretic paradigm to study honest behavior is the sender-receiver game  \cite{gneezy2005deception}. In this game, player 1 is given a private information (for example, the outcome of a die) and is asked to communicate this piece of information to player 2. Player 1 can either communicate the truthful piece of information, or can lie. The role of player 2 is to guess the original piece of information. If player 2 guesses the original piece of information, then players 1 and 2 are both paid according to some option A. Conversely, if player 2 does not guess the original piece of information, then players 1 and 2 are both paid according to option B. Crucially, only player 1 knows the payoffs associated with options A and B. A variant of this game in which player 2 makes no choice has also been introduced and studied \cite{biziou2015does, capraro2017does}, in order to avoid the confound of \emph{sophisticated deception}, that is, players who tell the truth because they believe that player 2 will not believe them \cite{sutter2009deception}.

Other important forms of moral behavior that ought to be investigated are `equity', that is, a desire to minimize payoff differences among players; `efficiency', that is, a desire to maximize the total welfare; and `maximin', that is, a desire to maximize the worse off payoff. These types of behavior are usually studied using simple distribution experiments, in which people have to decide between two or more allocations of money \cite{charness2002understanding, engelmann2004inequality, capraro2014benevolent, capraro2018right, tappin2018doing}.

\section*{Discussion}

Methods of statistical physics and network science have proven to be very valuable for successfully studying the evolution of cooperation in social dilemma games. However, empirical research shows that this kind of behavior is only one form of a more general class of moral behavior. The later includes love your family, help your group, return favors, be brave, defer to authority, be fair, respect others' property, honesty, equity, and efficiency, as well as many others. We have outlined a set of games and mathematical models that could be used efficiently to study particular aspects of some of these forms of moral behavior.

Taken together, the application of statistical physics to study the evolution of moral behavior has the potential to become a flourishing and vibrant avenue of future research. We believe so for two reasons. In the first place, it would allow us to understand why our societies evolved as they did and which moral principles are more likely to evolve. Secondly, since many social conflicts are ultimately conflicts between different moral positions \cite{nagel1987moral, pearce1997moral, bartos2002using}, exploring the evolution of moral behavior could allow us to predict in advance the consequences of a moral conflict, and suggest strategies to avoid it or, in case it is unavoidable, strategies to minimize its costs. We hope that at least parts of our vision will be put to practice in the near future.

\begin{acknowledgments}
This work was supported by the Slovenian Research Agency (Grants J1-7009, J4-9302, J1-9112 and P5-0027).
\end{acknowledgments}


\begin{thebibliography}{100}

\bibitem{pinker2011better}
Pinker, S.
\newblock {\em The better angels of our nature: The decline of violence in
  history and its causes}.
\newblock Penguin uk,  (2011).

\bibitem{pinker2018enlightenment}
Pinker, S.
\newblock {\em Enlightenment now: the case for reason, science, humanism, and
  progress}.
\newblock Penguin,  (2018).

\bibitem{ostrom2000collective}
Ostrom, E.
\newblock Collective action and the evolution of social norms.
\newblock {\em Journal of Economic Perspectives}{ \bf 14}, 137--158 (2000).

\bibitem{fehr2002altruistic}
Fehr, E. and G{\"a}chter, S.
\newblock Altruistic punishment in humans.
\newblock {\em Nature}{ \bf 415}, 137 (2002).

\bibitem{milinski2002reputation}
Milinski, M., Semmann, D., and Krambeck, H.-J.
\newblock Reputation helps solve the ?tragedy of the commons?
\newblock {\em Nature}{ \bf 415}, 424 (2002).

\bibitem{gintis2003explaining}
Gintis, H., Bowles, S., Boyd, R., and Fehr, E.
\newblock Explaining altruistic behavior in humans.
\newblock {\em Evolution and Human Behavior}{ \bf 24}, 153--172 (2003).

\bibitem{fehr2004social}
Fehr, E. and Fischbacher, U.
\newblock Social norms and human cooperation.
\newblock {\em Trends in Cognitive Sciences}{ \bf 8}, 185--190 (2004).

\bibitem{henrich2006costly}
Henrich, J., McElreath, R., Barr, A., Ensminger, J., Barrett, C., Bolyanatz,
  A., Cardenas, J.~C., Gurven, M., Gwako, E., Henrich, N., et~al.
\newblock Costly punishment across human societies.
\newblock {\em Science}{ \bf 312}, 1767--1770 (2006).

\bibitem{nowak2006five}
Nowak, M.~A.
\newblock Five rules for the evolution of cooperation.
\newblock {\em Science}{ \bf 314}, 1560--1563 (2006).

\bibitem{herrmann2008antisocial}
Herrmann, B., Th{\"o}ni, C., and G{\"a}chter, S.
\newblock Antisocial punishment across societies.
\newblock {\em Science}{ \bf 319}, 1362--1367 (2008).

\bibitem{bowles2011cooperative}
Bowles, S. and Gintis, H.
\newblock {\em A cooperative species: Human reciprocity and its evolution}.
\newblock Princeton University Press,  (2011).

\bibitem{capraro2013model}
Capraro, V.
\newblock A model of human cooperation in social dilemmas.
\newblock {\em PLoS ONE}{ \bf 8}, e72427 (2013).

\bibitem{rand2013human}
Rand, D.~G. and Nowak, M.~A.
\newblock Human cooperation.
\newblock {\em Trends in Cognitive Sciences}{ \bf 17}, 413--425 (2013).

\bibitem{perc2017statistical}
Perc, M., Jordan, J.~J., Rand, D.~G., Wang, Z., Boccaletti, S., and Szolnoki,
  A.
\newblock Statistical physics of human cooperation.
\newblock {\em Physics Reports}{ \bf 687}, 1--51 (2017).

\bibitem{tomasello2005search}
Tomasello, M., Carpenter, M., Call, J., Behne, T., and Moll, H.
\newblock In search of the uniquely human.
\newblock {\em Behavioral and Brain Sciences}{ \bf 28}, 721--727 (2005).

\bibitem{hrdy_11}
Hrdy, S.~B.
\newblock {\em Mothers and Others: The Evolutionary Origins of Mutual
  Understanding}.
\newblock Harvard University Press, Cambridge, MA,  (2011).

\bibitem{nowak_11}
Nowak, M.~A. and Highfield, R.
\newblock {\em SuperCooperators: Altruism, Evolution, and Why We Need Each
  Other to Succeed}.
\newblock Free Press, New York,  (2011).

\bibitem{capraro2018right}
Capraro, V. and Rand, D.~G.
\newblock Do the right thing: Experimental evidence that preferences for moral
  behavior, rather than equity or efficiency per se, drive human prosociality.
\newblock {\em Judgment and Decision Making}{ \bf 13}, 99--111 (2018).

\bibitem{rapoport1965prisoner}
Rapoport, A. and Chammah, A.~M.
\newblock {\em Prisoner's dilemma: A study in conflict and cooperation}, volume
  165.
\newblock University of Michigan Press,  (1965).

\bibitem{skyrms_04}
Skyrms, B.
\newblock {\em The Stag Hunt and the Evolution of Social Structure}.
\newblock Cambridge University Press, Cambridge, U.K.,  (2004).

\bibitem{hardin2009tragedy}
Hardin, G.
\newblock The tragedy of the commons.
\newblock {\em Journal of Natural Resources Policy Research}{ \bf 1},
  243--253 (2009).

\bibitem{hamilton1964genetical}
Hamilton, W.~D.
\newblock The genetical evolution of social behaviour. i.
\newblock {\em Journal of Theoretical Biology}{ \bf 7}, 1--16 (1964).

\bibitem{trivers1971evolution}
Trivers, R.~L.
\newblock The evolution of reciprocal altruism.
\newblock {\em The Quarterly Review of Biology}{ \bf 46}, 35--57 (1971).

\bibitem{nowak1998evolution}
Nowak, M.~A. and Sigmund, K.
\newblock Evolution of indirect reciprocity by image scoring.
\newblock {\em Nature}{ \bf 393}, 573 (1998).

\bibitem{fehr1999theory}
Fehr, E. and Schmidt, K.~M.
\newblock A theory of fairness, competition, and cooperation.
\newblock {\em The Quarterly Journal of Economics}{ \bf 114}, 817--868
  (1999).

\bibitem{bolton2000erc}
Bolton, G.~E. and Ockenfels, A.
\newblock Erc: A theory of equity, reciprocity, and competition.
\newblock {\em The American Economic Review}{ \bf 90}, 166--193 (2000).

\bibitem{charness2002understanding}
Charness, G. and Rabin, M.
\newblock Understanding social preferences with simple tests.
\newblock {\em The Quarterly Journal of Economics}{ \bf 117}, 817--869
  (2002).

\bibitem{engelmann2004inequality}
Engelmann, D. and Strobel, M.
\newblock Inequality aversion, efficiency, and maximin preferences in simple
  distribution experiments.
\newblock {\em The American Economic Review}{ \bf 94}, 857--869 (2004).

\bibitem{rand2012spontaneous}
Rand, D.~G., Greene, J.~D., and Nowak, M.~A.
\newblock Spontaneous giving and calculated greed.
\newblock {\em Nature}{ \bf 489}, 427 (2012).

\bibitem{capraro2015translucent}
Capraro, V. and Halpern, J.~Y.
\newblock Translucent players: {E}xplaining cooperative behavior in social
  dilemmas.
\newblock {\em Proceedings of the 15th conference on Theoretical Aspects of
  Rationality and Knowledge}{ \bf 215}, 114--126 (2016).

\bibitem{halpern2010cooperative}
Halpern, J.~Y. and Rong, N.
\newblock Cooperative equilibrium.
\newblock In {\em Proceedings of the 9th International Conference on Autonomous
  Agents and Multiagent Systems: volume 1-Volume 1},  1465--1466. International
  Foundation for Autonomous Agents and Multiagent Systems, (2010).

\bibitem{barcelo2015group}
Barcelo, H. and Capraro, V.
\newblock Group size effect on cooperation in one-shot social dilemmas.
\newblock {\em Scientific Reports}{ \bf 5}, 7937 (2015).

\bibitem{rand_pnas14}
Rand, D.~G., Nowak, M.~A., Fowler, J.~H., and Christakis, N.~A.
\newblock Static network structure can stabilize human cooperation.
\newblock {\em Proc. Natl. Acad. Sci. USA}{ \bf 111}, 17093--17098 (2014).

\bibitem{nowak1992evolutionary}
Nowak, M.~A. and May, R.~M.
\newblock Evolutionary games and spatial chaos.
\newblock {\em Nature}{ \bf 359}, 826 (1992).

\bibitem{lieberman2005evolutionary}
Lieberman, E., Hauert, C., and Nowak, M.~A.
\newblock Evolutionary dynamics on graphs.
\newblock {\em Nature}{ \bf 433}, 312 (2005).

\bibitem{ohtsuki2006simple}
Ohtsuki, H., Hauert, C., Lieberman, E., and Nowak, M.~A.
\newblock A simple rule for the evolution of cooperation on graphs and social
  networks.
\newblock {\em Nature}{ \bf 441}, 502 (2006).

\bibitem{perc_bs10}
Perc, M. and Szolnoki, A.
\newblock {Coevolutionary games -- A mini review}.
\newblock {\em BioSystems}{ \bf 99}, 109--125 (2010).

\bibitem{wang_epjb15}
Wang, Z., Wang, L., Szolnoki, A., and Perc, M.
\newblock {Evolutionary games on multilayer networks: A colloquium}.
\newblock {\em European Physical Journal B}{ \bf 88}, 124 (2015).

\bibitem{santos_prl05}
Santos, F.~C. and Pacheco, J.~M.
\newblock Scale-free networks provide a unifying framework for the emergence of
  cooperation.
\newblock {\em Phys. Rev. Lett.}{ \bf 95}, 098104 (2005).

\bibitem{pacheco_prl06}
Pacheco, J.~M., Traulsen, A., and Nowak, M.~A.
\newblock Coevolution of strategy and structure in complex networks with
  dynamical linking.
\newblock {\em Phys. Rev. Lett.}{ \bf 97}, 258103 (2006).

\bibitem{gomez-gardenes_prl07}
G{\'o}mez-Garde{\~n}es, J., Campillo, M., Flor{\'{\i}}a, L.~M., and Moreno, Y.
\newblock Dynamical organization of cooperation in complex networks.
\newblock {\em Phys. Rev. Lett.}{ \bf 98}, 108103 (2007).

\bibitem{ohtsuki_prl07}
Ohtsuki, H., Nowak, M.~A., and Pacheco, J.~M.
\newblock Breaking the symmetry between interaction and replacement in
  evolutionary dynamics on graphs.
\newblock {\em Phys. Rev. Lett.}{ \bf 98}, 108106 (2007).

\bibitem{lee_s_prl11}
Lee, S., Holme, P., and Wu, Z.-X.
\newblock Emergent hierarchical structures in multiadaptive games.
\newblock {\em Phys. Rev. Lett.}{ \bf 106}, 028702 (2011).

\bibitem{mathiesen_prl11}
Mathiesen, J., Mitarai, N., Sneppen, K., and Trusina, A.
\newblock Ecosystems with mutually exclusive interactions self-organize to a
  state of high diversity.
\newblock {\em Phys. Rev. Lett.}{ \bf 107}, 188101 (2011).

\bibitem{szolnoki2012defense}
Szolnoki, A., Perc, M., and Szabo, G.
\newblock Defense mechanisms of empathetic players in the spatial ultimatum
  game.
\newblock {\em Physical Review Letters}{ \bf 109}, 078701 (2012).

\bibitem{assaf_prl12}
Assaf, M. and Mobilia, M.
\newblock Metastability and anomalous fixation in evolutionary games on
  scale-free networks.
\newblock {\em Phys. Rev. Lett.}{ \bf 109}, 188701 (2012).

\bibitem{gomez_prl13}
G{\'o}mez, S., D{\'i}az-Guilera, A., G{\'o}mez-Garde{\~n}es, J.,
  P{\'e}rez-Vicente, C., Moreno, Y., and Arenas, A.
\newblock Diffusion dynamics on multiplex networks.
\newblock {\em Phys. Rev. Lett.}{ \bf 110}, 028701 (2013).

\bibitem{knebel_prl13}
Knebel, J., Kr{\"u}ger, T., Weber, M., and Frey, E.
\newblock Coexistence and survival in conservative \protect{Lotka-Volterra}
  networks.
\newblock {\em Phys. Rev. Lett.}{ \bf 110}, 168106 (2013).

\bibitem{pinheiro_prl14}
Pinheiro, F., Santos, M.~D., Santos, F., and Pacheco, J.
\newblock Origin of peer influence in social networks.
\newblock {\em Phys. Rev. Lett.}{ \bf 112}, 098702 (2014).

\bibitem{szolnoki_pre09c}
Szolnoki, A., Perc, M., and Szab{\'o}, G.
\newblock {Topology-independent impact of noise on cooperation in spatial
  public goods games}.
\newblock {\em Physical Review E}{ \bf 80}, 056109 (2009).

\bibitem{perc_ejp17}
Perc, M.
\newblock {High-performance parallel computing in the classroom using the
  public goods game as an example}.
\newblock {\em European Journal of Physics}{ \bf 38}, 045801 (2017).

\bibitem{helbing_njp10}
Helbing, D., Szolnoki, A., Perc, M., and Szab{\'o}, G.
\newblock {Punish, but not too hard: How costly punishment spreads in the
  spatial public goods game}.
\newblock {\em New Journal of Physics}{ \bf 12}, 083005 (2010).

\bibitem{szolnoki_pre11a}
Szolnoki, A., Szab{\'o}, G., and Perc, M.
\newblock {Phase diagrams for the spatial public goods game with pool
  punishment}.
\newblock {\em Physical Review E}{ \bf 83}, 036101 (2011).

\bibitem{perc_njp12}
Perc, M. and Szolnoki, A.
\newblock {Self-organization of punishment in structured populations}.
\newblock {\em New Journal of Physics}{ \bf 14}, 043013 (2012).

\bibitem{wang2013impact}
Wang, Z., Xia, C.-Y., Meloni, S., Zhou, C.-S., and Moreno, Y.
\newblock Impact of social punishment on cooperative behavior in complex
  networks.
\newblock {\em Scientific Reports}{ \bf 3}, 3055 (2013).

\bibitem{chen_njp14}
Chen, X., Szolnoki, A., and Perc, M.
\newblock {Probabilistic sharing solves the problem of costly punishment}.
\newblock {\em New Journal of Physics}{ \bf 16}, 083016 (2014).

\bibitem{chen_pre15}
Chen, X., Szolnoki, A., and Perc, M.
\newblock {Competition and cooperation among different punishing strategies in
  the spatial public goods game}.
\newblock {\em Physical Review E}{ \bf 92}, 012819 (2015).

\bibitem{szolnoki_prx17}
Szolnoki, A. and Perc, M.
\newblock {Second-order free-riding on antisocial punishment restores the
  effectiveness of prosocial punishment}.
\newblock {\em Physical Review X}{ \bf 7}, 041027 (2017).

\bibitem{szolnoki_epl10}
Szolnoki, A. and Perc, M.
\newblock {Reward and cooperation in the spatial public goods game}.
\newblock {\em EPL (Europhysics Letters)}{ \bf 92}, 38003 (2010).

\bibitem{hilbe_prsb10}
Hilbe, C. and Sigmund, K.
\newblock Incentives and opportunism: from the carrot to the stick.
\newblock {\em Proc. R. Soc. B}{ \bf 277}, 2427--2433 (2010).

\bibitem{szolnoki_njp12}
Szolnoki, A. and Perc, M.
\newblock {Evolutionary advantages of adaptive rewarding}.
\newblock {\em New Journal of Physics}{ \bf 14}, 093016 (2012).

\bibitem{szolnoki_prsb15}
Szolnoki, A. and Perc, M.
\newblock {Antisocial pool rewarding does not deter public cooperation}.
\newblock {\em Proceedings of the Royal Society B}{ \bf 282}, 20151975 (2015).

\bibitem{szolnoki_njp16}
Szolnoki, A. and Perc, M.
\newblock {Competition of tolerant strategies in the spatial public goods
  game}.
\newblock {\em New Journal of Physics}{ \bf 18}, 083021 (2016).

\bibitem{perc_ejp18}
Perc, M.
\newblock {Stability of subsystem solutions in agent-based models}.
\newblock {\em European Journal of Physics}{ \bf 39}, 014001 (2018).

\bibitem{dickman_pre01}
Dickman, R.
\newblock First- and second-order phase transitions in a driven lattice gas
  with nearest-neighbor exclusion.
\newblock {\em Phys. Rev. E}{ \bf 64}, 016124 (2001).

\bibitem{szolnoki_pre02}
Szolnoki, A.
\newblock Dynamical mean-field approximation for a pair contact process with a
  particle source.
\newblock {\em Phys. Rev. E}{ \bf 66}, 057102 (2002).

\bibitem{dickman_pre02}
Dickman, R.
\newblock \protect{\textit{n}-site} approximations and coherent-anomaly-method
  analysis for a stochastic sandpile.
\newblock {\em Phys. Rev. E}{ \bf 66}, 036122 (2002).

\bibitem{helbing_ploscb10}
Helbing, D., Szolnoki, A., Perc, M., and Szab{\'o}, G.
\newblock {Evolutionary establishment of moral and double moral standards
  through spatial interactions}.
\newblock {\em PLoS Computational Biology}{ \bf 6}, e1000758 (2010).

\bibitem{szolnoki_prx13}
Szolnoki, A. and Perc, M.
\newblock {Correlation of positive and negative reciprocity fails to confer an
  evolutionary advantage: Phase transitions to elementary strategies}.
\newblock {\em Physical Review X}{ \bf 3}, 041021 (2013).

\bibitem{greene2015rise}
Greene, J.~D.
\newblock The rise of moral cognition.
\newblock {\em Cognition}{ \bf 135}, 39--42 (2015).

\bibitem{tomasello2013origins}
Tomasello, M. and Vaish, A.
\newblock Origins of human cooperation and morality.
\newblock {\em Annual review of psychology}{ \bf 64}, 231--255 (2013).

\bibitem{rawls2009theory}
Rawls, J.
\newblock {\em A theory of justice: Revised edition}.
\newblock Harvard university press,  (2009).

\bibitem{mackie1990ethics}
Mackie, J.
\newblock {\em Ethics: Inventing right and wrong}.
\newblock Penguin UK,  (1990).

\bibitem{wong1984moral}
Wong, D.~B.
\newblock {\em Moral relativity}.
\newblock Univ of California Press,  (1984).

\bibitem{rai2011moral}
Rai, T.~S. and Fiske, A.~P.
\newblock Moral psychology is relationship regulation: moral motives for unity,
  hierarchy, equality, and proportionality.
\newblock {\em Psychological Review}{ \bf 118}, 57 (2011).

\bibitem{curry2016morality}
Curry, O.~S.
\newblock Morality as cooperation: A problem-centred approach.
\newblock In {\em The evolution of morality},  27--51. Springer (2016).

\bibitem{curry2018good}
Curry, O., Mullins, D., and Whitehouse, H.
\newblock Is it good to cooperate? testing the theory of
  morality-as-cooperation in 60 societies.
\newblock {\em Current Anthropology}{ \bf 1}, osf.io/9546r (2018).

\bibitem{berg1995trust}
Berg, J., Dickhaut, J., and McCabe, K.
\newblock Trust, reciprocity, and social history.
\newblock {\em Games and Economic Behavior}{ \bf 10}, 122--142 (1995).

\bibitem{tajfel1970experiments}
Tajfel, H.
\newblock Experiments in intergroup discrimination.
\newblock {\em Scientific American}{ \bf 223}, 96--103 (1970).

\bibitem{guth1982experimental}
G{\"u}th, W., Schmittberger, R., and Schwarze, B.
\newblock An experimental analysis of ultimatum bargaining.
\newblock {\em Journal of Economic Behavior \& Organization}{ \bf 3},
  367--388 (1982).

\bibitem{page_prsb00}
Page, K.~M., Nowak, M.~A., and Sigmund, K.
\newblock The spatial ultimatum game.
\newblock {\em Proc. R. Soc. Lond. B}{ \bf 267}, 2177--2182 (2000).

\bibitem{kuperman_epjb08}
Kuperman, M.~N. and Risau-Gusman, S.
\newblock The effect of topology on the spatial ultimatum game.
\newblock {\em Eur. Phys. J. B}{ \bf 62}, 233--238 (2008).

\bibitem{eguiluz_acs09}
Equ{\'{\i}}luz, V.~M. and Tessone, C.
\newblock Critical behavior in an evolutionary ultimatum game with social
  structure.
\newblock {\em Adv. Complex Systems}{ \bf 12}, 221--232 (2009).

\bibitem{da-silva_r_jtb09}
da~Silva, R., Kellerman, G.~A., and Lamb, L.~C.
\newblock Statistical fluctuations in population bargaining in the ultimatum
  game: Static and evolutionary aspects.
\newblock {\em J. Theor. Biol.}{ \bf 258}, 208--218 (2009).

\bibitem{deng_ll_pa11}
Deng, L., Tang, W., and Zhang, J.
\newblock The coevolutionay ultimatum game on different network topologies.
\newblock {\em Physica A}{ \bf 390}, 4227--4235 (2011).

\bibitem{gao_j_epl11}
Gao, J., Li, Z., Wu, T., and Wang, L.
\newblock The coevolutionary ultimatum game.
\newblock {\em EPL}{ \bf 93}, 48003 (2011).

\bibitem{szolnoki2012accuracy}
Szolnoki, A., Perc, M., and Szab{\'o}, G.
\newblock Accuracy in strategy imitations promotes the evolution of fairness in
  the spatial ultimatum game.
\newblock {\em EPL (Europhysics Letters)}{ \bf 100}, 28005 (2012).

\bibitem{deng_ll_jsm12}
Deng, L., Wang, C., Tang, W., Zhou, G., and Cai, J.
\newblock A network growth model based on the evolutionary ultimatum game.
\newblock {\em J. Stat. Mech.}{ \bf 2012}, P11013 (2012).

\bibitem{iranzo_pone12}
Iranzo, J., Flor{\'{\i}}a, L., Moreno, Y., and S{\'a}nchez, A.
\newblock Empathy emerges spontaneously in the ultimatum game: Small groups and
  networks.
\newblock {\em PLoS ONE}{ \bf 7}, e43781 (2011).

\bibitem{miyaji_csf13}
Miyaji, K., Wang, Z., Tanimoto, J., Hagishima, A., and Kokubo, S.
\newblock The evolution of fairness in the coevolutionary ultimatum games.
\newblock {\em Chaos, Solitons \& Fractals}{ \bf 56}, 13--18 (2013).

\bibitem{forsythe1994fairness}
Forsythe, R., Horowitz, J.~L., Savin, N.~E., and Sefton, M.
\newblock Fairness in simple bargaining experiments.
\newblock {\em Games and Economic Behavior}{ \bf 6}, 347--369 (1994).

\bibitem{krupka2013identifying}
Krupka, E.~L. and Weber, R.~A.
\newblock Identifying social norms using coordination games: Why does dictator
  game sharing vary?
\newblock {\em Journal of the European Economic Association}{ \bf 11},
  495--524 (2013).

\bibitem{gneezy2005deception}
Gneezy, U.
\newblock Deception: The role of consequences.
\newblock {\em The American Economic Review}{ \bf 95}, 384--394 (2005).

\bibitem{biziou2015does}
Biziou-van Pol, L., Haenen, J., Novaro, A., Liberman, A.~O., and Capraro, V.
\newblock Does telling white lies signal pro-social preferences?
\newblock {\em Judgment and Decision Making}{ \bf 10}, 538--548 (2015).

\bibitem{capraro2017does}
Capraro, V.
\newblock Does the truth come naturally? time pressure increases honesty in
  one-shot deception games.
\newblock {\em Economics Letters}{ \bf 158}, 54--57 (2017).

\bibitem{sutter2009deception}
Sutter, M.
\newblock Deception through telling the truth?! experimental evidence from
  individuals and teams.
\newblock {\em The Economic Journal}{ \bf 119}, 47--60 (2009).

\bibitem{capraro2014benevolent}
Capraro, V., Smyth, C., Mylona, K., and Niblo, G.~A.
\newblock Benevolent characteristics promote cooperative behaviour among
  humans.
\newblock {\em PLoS ONE}{ \bf 9}, e102881 (2014).

\bibitem{tappin2018doing}
Tappin, B.~M. and Capraro, V.
\newblock Doing good vs. avoiding bad in prosocial choice: A refined test and
  extension of the morality preference hypothesis.
\newblock {\em Journal of Experimental Social Psychology}{ \bf 79}, 64--70
  (2018).

\bibitem{nagel1987moral}
Nagel, T.
\newblock Moral conflict and political legitimacy.
\newblock {\em Philosophy \& Public Affairs}{ \bf 16}, 215--240 (1987).

\bibitem{pearce1997moral}
Pearce, W.~B. and Littlejohn, S.~W.
\newblock {\em Moral conflict: When social worlds collide}.
\newblock Sage,  (1997).

\bibitem{bartos2002using}
Bartos, O.~J. and Wehr, P.
\newblock {\em Using conflict theory}.
\newblock Cambridge University Press,  (2002).

\end{thebibliography}
\end{document}